\documentclass{article}
\usepackage{amssymb}
\usepackage{amsmath}

\input{tcilatex}
\begin{document}

\begin{center}
{\LARGE Isochronous Spacetimes}

\bigskip

$^{\ast \heartsuit }$\textbf{Fabio Briscese}$^{1}$ and $^{+\lozenge }$%
\textbf{Francesco Calogero}$^{2}\bigskip $

$^{\ast }$Istituto Nazionale di Alta Matematica Francesco Severi, Gruppo
Nazionale di Fisica Matematica, Citt\`{a} Universitaria, P.le A. Moro 5,
00185 Rome, Italy

$^{\heartsuit }$Dipartimento SBAI, Sezione di Matematica, University of Rome
\textquotedblleft La Sapienza", via Antonio Scarpa 16, 00161 Roma, Italy

$^{+}$Physics Department, University of Rome \textquotedblleft La Sapienza",
p. Aldo Moro, I-00185 ROMA, Italy

$^{\lozenge }$Istituto Nazionale di Fisica Nucleare, Sezione di Roma

$^{1}$fabio.briscese@sbai.uniroma1.it, briscese.phys@gmail.com

$^{2}$francesco.calogero@roma1.infn.it, francesco.calogero@uniroma1.it

\bigskip

\textit{Abstract}
\end{center}

The possibility has been recently demonstrated to manufacture
(nonrelativistic, Hamiltonian) many-body problems which feature an \textit{%
isochronous} time evolution with an arbitrarily assigned period $T$ yet
mimic with good approximation, or even exactly, any given many-body problem
(within a quite large class, encompassing most of nonrelativistic physics)
over times $\tilde{T}$ which may also be arbitrarily large (but of course
such that $\tilde{T}<T$). In this paper we review and further explore the
possibility to extend this finding to a general relativity context, so that
it becomes relevant for cosmology.

\section{Introduction}

In this paper we revisit the findings reported in \cite{isochronous
cosmologies} and report some additional considerations relevant for a better
understanding of the validity---in the context of theoretical and
mathematical physics---of those results.

It has been recently shown \cite{CL2007,calogero} that---given a general
autonomous dynamical system $D$, other autonomous dynamical systems $\tilde{D%
}$ can be manufactured, featuring two additional \textit{arbitrary} positive
parameters $T$ and $\tilde{T}$ with $T>\tilde{T}$ and possibly also two
additional dynamical variables---which are characterized by the following
two properties:

(i) For the same variables of the original dynamical system $D$ the new
dynamical system $\tilde{D}$ yields, over the time interval $\tilde{T}$,
hence over an \textit{arbitrarily long} time, a dynamical evolution which
mimics \textit{arbitrarily closely} that yielded by the original system $D$;
up to corrections of order $t/\tilde{T}$, or possibly even \textit{%
identically}.

(ii) The system $\tilde{D}$ is \textit{isochronous}: for arbitrary initial
data all its solutions are completely periodic with a period $T$, which can
also be arbitrarily assigned, except for the\ (obviously necessary)
condition $T>\tilde{T}$.

Moreover it has been shown \cite{CL2007,calogero} that, if the dynamical
system $D$ is a many-body problem characterized by a (standard, autonomous)
Hamiltonian $H$ which is translation-invariant (i. e., it features no
external forces), other (also autonomous) Hamiltonians $\tilde{H}$
characterizing \textit{modified} many-body problems can be manufactured
which feature the \textit{same} dynamical variables as $H$ (i. e., in this
case there is no need to introduce two additional dynamical variables) and
which yield a time evolution \textit{quite close}, or even \textit{identical}%
, to that yielded by the original Hamiltonian $H$ over the \textit{%
arbitrarily assigned} time $\tilde{T}$, while being \textit{isochronous}
with the \textit{arbitrarily assigned} period $T$, of course with $T>\tilde{T%
}$.

Let us emphasize that the class of Hamiltonians $H$ for which this result is
valid is quite general. In particular it includes the standard Hamiltonian
system describing an \textit{arbitrary} number $N$ of point particles with
\textit{arbitrary} masses moving in a space of \textit{arbitrary} dimensions
$d$ and interacting among themselves via potentials depending \textit{%
arbitrarily} from the interparticle distances (including the possibility of
multiparticle forces), being therefore generally valid for any realistic
many-body problem, hence encompassing most of nonrelativistic physics. This
result is moreover true, \textit{mutatis mutandis}, in a quantal context.

For instance let us tersely review here the case of the standard Hamiltonian
describing the many-body problem, but focusing---merely for notational
simplicity---on the case with equal particles (and setting their mass to
unity) and a one-dimensional setting, which reads (in self-evident standard
notation)
\begin{subequations}
\begin{equation}
H\left( \underline{p},\underline{q}\right) =\frac{1}{2}\sum_{n=1}^{N}\left(
p_{n}^{2}\right) +V\left( \underline{q}\right)  \label{H}
\end{equation}%
with the potential being translation-invariant, $V\left( \underline{q}%
+a\right) =V\left( \underline{q}\right) $ (i. e., no external forces) but
otherwise unrestricted (except for the condition---again, for
simplicity---that the time-evolution yielded by this Hamiltonian be
nonsingular). The standard approach to treat this problem is to introduce
the center-of-mass coordinate $Q\left( \underline{q}\right) =\frac{1}{N}%
\sum_{n=1}^{N}q_{n}$ and the total momentum $P\left( \underline{p}\right)
=\sum_{n=1}^{N}p_{n},$ and to then focus on the coordinate and momenta in
the center of mass, $x_{n}=q_{n}-Q\left( \underline{q}\right) ,$ $%
y_{n}=p_{n}-P\left( \underline{p}\right) /N$, which characterize the physics
of the problem and whose evolution is determined by the Hamiltonian $h\left(
\underline{y},\underline{x}\right) $ defined as follows:%
\begin{equation}
H\left( \underline{p},\underline{q}\right) =\frac{\left[ P\left( \underline{p%
}\right) \right] ^{2}}{2N}+h\left( \underline{y},\underline{x}\right) ~,
\label{Hh}
\end{equation}%
\begin{equation}
h\left( \underline{y},\underline{x}\right) =\frac{1}{2}\sum_{n=1}^{N}\left(
y_{n}^{2}\right) +V\left( \underline{x}\right) ~,  \label{h}
\end{equation}%
where $V\left( \underline{x}\right) =V\left( \underline{q}\right) $ thanks
to the assumed translation-invariance of $V\left( \underline{q}\right) $. An
\textit{isochronous} Hamiltonian $\tilde{H}\left( \underline{p},\underline{q}%
;T\right) $ reads then as follows,
\end{subequations}
\begin{subequations}
\begin{equation}
\tilde{H}\left( \underline{p},\underline{q};T\right) =\frac{1}{2}%
\sum_{n=1}^{N}\left\{ \left[ P\left( \underline{p}\right) +h\left(
\underline{y},\underline{x}\right) \right] ^{2}+\left( \frac{2\pi }{T}%
\right) \left[ Q\left( \underline{q}\right) \right] ^{2}\right\} ~.
\label{Htilde}
\end{equation}%
It can indeed be shown \cite{CL2007,calogero} that it entails an \textit{%
isochronous} evolution (with period $T$) of the center of mass coordinate $Q$%
, of the total momentum $P$, and---most importantly---of all the particle
coordinates, whose evolution then reads%
\begin{equation}
\underline{\tilde{x}}\left( t\right) =\underline{x}\left( \tau \left(
t\right) \right) ~,~~~\underline{\tilde{y}}\left( t\right) =\underline{y}%
\left( \tau \left( t\right) \right) ~,  \label{xytildet}
\end{equation}%
where we indicate as $\underline{\tilde{x}}\left( t\right) ,$ $\underline{%
\tilde{y}}\left( t\right) $ the time evolution yielded by the tilded
Hamiltonian $\tilde{H}\left( \underline{p},\underline{q};T\right) $ and as $%
\underline{x}\left( t\right) ,$ $\underline{y}\left( t\right) $ the time
evolution yielded by the original Hamiltonian $H\left( \underline{p},%
\underline{q}\right) ,$ and%
\begin{equation}
\tau \left( t\right) =A~\sin \left( \frac{2\pi t}{T}+\Phi \right)
\label{taut}
\end{equation}%
where $A$ and $\Phi $ are \textit{constant} parameters given by simple
expressions in terms of the initial values, $Q\left( 0\right) $ and $P\left(
0\right) $, of the position of the center of mass of the system and of its
total momentum, and of the Hamiltonian $\tilde{H}\left( \underline{p},%
\underline{q};T\right) $ (which is of course a constant of motion for this
time evolution). Note that, for $\left\vert t\right\vert <<T,$%
\begin{equation}
\tau \left( t\right) =\alpha +\beta t+O\left[ \left( \frac{t}{T}\right) ^{2}%
\right] ~~~\text{with \hspace{0in} }\alpha =A~\sin \left( \Phi \right)
~,~~~\beta =\frac{2\pi A}{T}\cos \left( \Phi \right) ~;
\end{equation}%
hence up to small corrections (\textit{arbitrarily small} for an \textit{%
arbitrarily large} assignment of the period $T$ as long as $\left\vert
t\right\vert $ is in an assigned interval $\tilde{T}$, of course such that $%
\tilde{T}<<T$) the time evolution of the tilded coordinates approximates
that of the untilded coordinates, up to a constant shift and rescaling of
time.

Note that in this case---again, for the sake of simplicity---the modified,
\textit{isochronous} Hamiltonian $\tilde{H}\left( \underline{p},\underline{q}%
;T\right) $ features only one additional constant parameter, $T$; but it has
been shown \cite{CL2007,calogero} that an analogous treatment---involving a
somewhat more complicated definition of a modified, \textit{isochronous}
Hamiltonian, featuring then two arbitrarily assigned parameters, $T$ and $%
\tilde{T}<T$---allows to manufacture a modified Hamiltonian that reproduces
\textit{exactly} (up to a constant rescaling and shift of the time variable)
the same time evolution yielded by the original Hamiltonian over the
interval $\tilde{T}$ but is \textit{isochronous} with period $T$---and let
us again emphasize that this can be done for a many-body problem involving
an \textit{arbitrary} number of (possibly different) particles, moving in a
space of \textit{arbitrary} dimension $d$ and interacting via \textit{%
arbitrary} interparticle forces. Which therefore includes most of
(nonrelativistic) physics.

Since it is impossible to distinguish experimentally dynamical systems that
behave \textit{arbitrarily closely}, or even \textit{identically}, over an
\textit{arbitrarily long} period of time, this finding has various
remarkable implications in the context of theoretical and mathematical
physics. In particular it raises \cite{CL2007,calogero} interesting
questions about the distinction between integrable and nonintegrable
evolutions, about the definition of chaotic behavior, and---for macroscopic
systems featuring, say, a number of particles of the order of Avogadro's
number---about statistical mechanics and the second principle of
thermodynamics. And for, say, $10^{85}$ particles it seems to have some
cosmological relevance.

But the proper setting of cosmological theories is general relativity. For
this reason in our previous paper \cite{isochronous cosmologies} we
investigated whether and how this result could be extended to that context.
We concluded that this is possible provided one extends general relativity
by allowing \textit{degenerate} (i.e. non invertible) metrics. In particular
we have shown \cite{isochronous cosmologies} that for any homogeneous,
isotropic and spatially flat metric $g_{\mu \nu }$ satisfying Einstein's
equations and providing a model of the universe, it is possible to find a
different (also homogeneous, isotropic and spatially flat) metric solution $%
\tilde{g}_{\mu \nu }$ which is \textit{locally} (in time) diffeomorphic to $%
g_{\mu \nu }$---hence yields the same cosmology as $g_{\mu \nu }$ for any
observation over an \textit{arbitrary} time interval $\tilde{T}$---and is
cyclic\footnote{%
We prefer to talk of cyclic instead of periodic solutions due to the fact
that periodicity is not an invariant concept, since it is not invariant
under a redefinition of time. For a review of cyclic models see \cite{cyclic}%
.} (in fact periodic with an \textit{arbitrary} period $T>\tilde{T}$ in the
time coordinate $t\,$); but it is \textit{degenerate} at an infinite,
discrete sequence of times $t_{n}=t_{0}\pm nT/2,$ $n=0,1,2,...$ . We
interpreted these metrics as corresponding to \textit{isochronous
cosmologies }\cite{isochronous cosmologies}.

Let us also recall \cite{isochronous cosmologies} that this result is not
restricted to homogeneous, isotropic and spatially flat metrics: it can be
easily extended to any synchronous metric, therefore it is quite general
since most metrics can be written in synchronous form by a diffeomorphic
change of coordinates.

Due to the diffeomorphic correspondence between $g_{\mu \nu }$ and $\tilde{g}%
_{\mu \nu }$ \textit{locally} in time---for time intervals of order $\tilde{T%
}$---these two metrics give the same physics locally in time, and therefore
there is no way to distinguish them using observations local in time; which
is essentially the same finding valid in the context of the Hamiltonian
systems considered in \cite{CL2007,calogero} and tersely recalled above. In
particular, the metric $\tilde{g}_{\mu \nu }$---while being cyclic on time
scales larger than $\tilde{T}$---may yield over the time interval $\tilde{T}$
just the standard sequence of domination firstly by radiation, then by
matter and dark energy, as well as the cosmological perturbations consistent
with all observational tests \cite%
{cmb,leansing,bao,lss,supernovae,ref0,ref1,ref2,ref3,ref4,bicep2}, which
characterize the metric $g_{\mu \nu }$ of the standard $\Lambda $-CDM
cosmological model (see for instance \cite{mukhanov} for a review).
Furthermore, it has been shown that such isochronous metrics $\tilde{g}_{\mu
\nu }$ can be manufactured to be geodesically complete \cite{isochronous
cosmologies} and therefore singularity-free\footnote{%
A spacetime is singularity-free if it is geodesically complete, i.e. if its
geodesics can be always past- and future-extended \cite{poisson,MTW}.}, so
that the geodesic motion as well as all physical quantities described by
scalar invariants are always well defined \cite{isochronous cosmologies} and
the Big Bang singularity may be avoided---even when some of the
phenomenological observations can nevertheless be interpreted as remnants of
a past Big Bang, which however may never be actually attained by the metric $%
\tilde{g}_{\mu \nu }$, neither in the past nor in the future. This implies
that these isochronous metrics $\tilde{g}_{\mu \nu }$ may describe a
singularity-free universe which---while featuring a time evolution which
reproduces identically (up to diffeomorphic time reparameterization; \textit{%
locally} in time, except at a discrete set of instants $t_{n}$) the standard
cosmological model characterized by the metric $g_{\mu \nu }$---features an
expansion which stops at some instant, to be followed by a period of
contraction, until this phase of evolution stops and is again followed by an
expanding phase, this pattern being repeated \textit{ad infinitum}. Let us
re-emphasize that the universe characterized by the metric $\tilde{g}_{\mu
\nu }$ may thereby avoid to experience the Big Bang singularity, even if the
universe which it mimics locally in time, characterized by the metric $%
g_{\mu \nu },$ does encounter that singularity. Let us also note that this
class of metrics realize \textit{de facto} the reversal of time's arrow, as
discussed for instance in \cite{sakharov}.

But in order to do so, these isochronous metrics must be \textit{degenerate}
at a discrete set of instants $t_{n}$ when the time reversals occur, say
from expansion to contraction and viceversa; a phenomenology whose inclusion
in general relativity might be considered problematic, because at these
instants $t_{n}$---when the metric is \textit{degenerate}---the "equivalence
principle" corresponding to the requirement that the metric tensor have a
Minkowskian signature is violated. On the other hand a \textit{physically
unobservable} violation of a "principle" can be hardly considered \textit{%
physically relevant}.

Purpose and scope of this paper is to better clarify the meaning and the
interpretation of the \textit{isochronous spacetimes} introduced in \cite%
{isochronous cosmologies}, and to discuss some additional technical aspects.
As stated above, these isochronous metrics have the property to be \textit{%
degenerate} at an infinite set of times $t_{n}$, i.e. over an infinite
number of $3$-dimensional hypersurfaces, where generally the scale factor
reaches its maximum and minimum values and the spacetime changes from an
expanding to a contracting phase and vice versa \cite{isochronous
cosmologies}.

The inclusion of degenerate metrics in general relativity is not a trivial
matter; indeed, it is forbidden if it were required---as an absolute (as it
were, "metaphysical") rule---that general relativity be consistent with the
"equivalence principle", i. e. that spacetime \textit{always} be a
Pseudo-Riemannian manifold \textit{with Minkowskian signature}. This would
indeed imply that the solutions of Einstein's general relativity, to be
acceptable, must be \textit{nondegenerate} metric tensors, since wherever in
spacetime a metric is \textit{degenerate} it is not possible to define its
inverse, hence the Christoffel symbols as well as the Riemann, Ricci and
Einstein tensors are not defined there. Hence---so the argument of some
critics goes---our isochronous metrics, which are not Minkowskian (locally,
on the hypersurfaces $t=t_{n}$) should be discarded in general relativity
because they violate the equivalence principle at those degeneracy surfaces.
Our rejoinder is that the metrics we introduced \cite{isochronous
cosmologies} are solutions of a version of general relativity whose
dynamics, while still governed by Einstein's equations\footnote{%
Einstein's equations by itself do not determine the signature of spacetime,
see the discussion in \cite{teitelboim,ellis1}.}, does allow the violation
of the equivalence principle at a discrete set of hypersurfaces. It can be
argued that such theories are therefore \textit{different} from standard
general relativity, if validity of the equivalence principle---enforced as
an absolute, universal rule---is considered an essential element of that
theory; although the difference is in fact \textit{physically unobservable}.
This situation is indeed quite analogous with the findings described above,
valid in the nonrelativistic context of Hamiltonian many-body problems
(classical or quantal), where the solutions of a physical theory described
by a Hamiltonian $H$ are arbitrarily well approximated or even identically
reproduced over an arbitrary time interval by periodic solutions of a
\textit{different} Hamiltonian $\tilde{H}$, i.e. by a \textit{different}
physical theory. The main difference is that in the general relativistic
context in\textbf{\ }the two \textit{different} physical theories the
dynamic is given by the\textit{\ same} Einstein's equations, only the class
of acceptable solutions is \textit{different}.

The consideration in the context of general relativity of degenerate metrics
is in any case not a novelty. For instance they were already introduced \cite%
{ellis1,ellis2}, and subsequently investigated in a series of papers \cite%
{elliscarfora,elliscomment,ellis3,ellis4,ellis5,ellis6,ellis8,ellis9,ellis10,ellis11,ellis12,ellis13,ellis14,ellis15,ellis16,ellis17,ellis18,ellis19}
, in order to investigate signature-changing spacetimes. This class of
signature-changing metrics corresponds to a classical realization of the
change of signature in quantum cosmology conjectured by Hartle and Hawking
\cite{hartlehawking3} (see also \cite%
{hartlehawking1,hartlehawking2,hartlehawking4}), which has philosophical
implications on the origin of the universe \cite{hawkinghistory}.

Below we will also consider a different realization of isochronous
cosmologies via non-degenerate metrics featuring a jump in their first
derivatives at the inversion times $t_{n}$, which then implies a
distributional contribution in the stress-energy tensor at $t_{n}$. These
metrics are diffeomorphic to the continuous and degenerate isochronous
metrics everywhere except on the inversion hypersurfaces $t=t_{n}$, hence
they describe the same physics except at the infinite set of discrete times $%
t_{n};$ therefore they may also be manufactured so as to agree with all
cosmological observations (\textit{locally} in time). Since these two
realizations of isochronous cosmologies are \textit{locally} but \textit{not
globally} diffeomorphic, they correspond to different spacetimes and may be
considered to emerge from \textit{different} theories: the nondegenerate
ones are generalized (in the sense of distributions) solutions of
Einsteinian general relativity (including universal validity of the
equivalence principle, but allowing jumps---whose physical significance and
justification is moot---at the discrete times $t_{n}$); the degenerate ones
feature no mathematical or physical pathologies but fail to satisfy the
"equivalence principle" at the discrete times $t_{n}$. They are essentially
indistinguishable---among themselves and from non-isochronous
cosmologies---by feasible experiments (unless one considers feasible an
experiment lasting an arbitrarily long time...).

\section{Isochronous Cosmologies}

The procedure used to construct, for an autonomous dynamical system $D$,
another autonomous dynamical system $\tilde{D}$ whose solutions \textit{%
approximate} arbitrarily closely, or even reproduce exactly, those of the
system $D$ over an \textit{arbitrary} time interval $\tilde{T}$ but are
\textit{isochronous} with an \textit{arbitrary} period $T>\tilde{T}$, is
based on the introduction of an auxiliary variable $\tau $ in place of the
physical time variable $t$ \cite{CL2007,calogero}. This procedure formally
entails a change in the time dependence of any physical quantity $f(t)$, so
that $f(t)$ gets replaced by $f(\tau \left( t\right) )$ with $\tau \left(
t\right) $ a periodic function of $t$ (with period $T$: $\tau \left( t\pm
T\right) =\tau \left( t\right) $), hence the new dynamics entails that all
physical quantities evolve periodically with period $T$. This change, in the
case of a nonrelativistic dynamical system, implies a change in the
dynamical equations: for instance, in the case of the (quite general)
many-body problem described above, the theory is characterized by a new
(autonomous) Hamiltonian $\tilde{H}$---different from the original
(autonomous) Hamiltonian $H$---which causes any physical variable originally
evolving as $f\left( t\right) $ under the dynamics yielded by $H$, to evolve
instead as $f(\tau (t))$ under the dynamics yielded by $\tilde{H}$. This
change of the physical laws determining the time evolution of the system
produces the \textit{isochronous} evolution \cite{calogero,CL2007}. But let
us stress---as we already did in \cite{isochronous cosmologies}---that any
attempt to attribute to the new variable $\tau $ the significance of "time"
would be improper and confusing: the variable playing the role of "time" is
the same, $t$, for both the original dynamical system $D$ and the modified
dynamical system $\tilde{D}$; in particular, for that characterized by the
standard many-body Hamiltonian $H,$ see (\ref{H}), as well as for that
characterized by the modified many-body Hamiltonian $\tilde{H}$, see (\ref%
{Htilde}).

The dynamical systems $D$ and $\tilde{D}$ mentioned above feature the same
trajectories in phase space, but while the time evolution of the dynamical
system $D$ corresponds to a \textit{uniform} forward motion along those
trajectories, the time evolutions of the modified dynamical systems $\tilde{D%
}$---although produced by time-independent equations of motion---correspond
to a \textit{periodic} (with assigned period $T$), forward and backward,
time evolution along those same trajectories, exploring therefore only a
portion of them; with the possibility to manufacture $\tilde{D}$ in such a
way that, for an \textit{arbitrary} subinterval $\tilde{T}<T,$ the dynamics
of $D$ and $\tilde{D}$ be \textit{very close} or even \textit{identical}.

In the context of general relativity, an analogous phenomenology exists \cite%
{isochronous cosmologies}. However, in this case the situation is a bit
different, due to the physical invariance of Einstein's equations under
diffeomorphisms, which implies that the procedure used to construct the
isochronous solutions, while again formally entailing the introduction of a
new auxiliary variable $\tau \left( t\right) $, does not imply any change of
the fundamental Einstein equations, corresponding instead to the
identification of an enlarged class of solutions of these equations. But
again any attempt to attribute \textit{globally} to this auxiliary variable $%
\tau $ the significance of \textit{time} would be improper and confusing;
while by attributing to it \textit{locally }(in time) the significance of
\textit{reparameterized time}, it is immediately seen that the enlarged
class of metrics entails dynamical evolutions which, \textit{locally in time}%
, are \textit{physically equivalent} to that yielded by the original,
unmodified metric.

Our point of departure is to assume a homogeneous, isotropic and spatially
flat metric $\ \tilde{g}_{\mu \nu }$ which, in a given reference frame $(t,%
\vec{x})$, is defined by the line element
\end{subequations}
\begin{equation}
ds^{2}=b(t)^{2}~dt^{2}-a(t)^{2}~d\vec{x}^{2}~,  \label{newmetric}
\end{equation}%
where $b\left( t\right) ^{2}$ is commonly termed the "lapse" function
(although we did not use this term in \cite{isochronous cosmologies}).

The lapse function is not determined by Einstein's equations; its
introduction in (\ref{newmetric}) corresponds to the freedom to
reparametrize time, which in general relativity is permissible, via a
diffeomorphic transformation, with no physical effect:\ indeed, in general
relativity it is the \textit{geometry} of spacetime that characterizes the
physics, and this geometry is not affected by the coordinates used to
describe it. An additional requirement which is generally considered
essential---amounting to the "equivalence principle"---is that the metric
have a \textit{Lorentzian signature} (everywhere in spacetime). This implies
that, in the case of the metric (\ref{newmetric}), the lapse function $b(t)$
must never vanish, so that one can always put this metric in its synchronous
form via the reparameterization of the time variable from $t$ to $\tau $
such that
\begin{equation}
d\tau =b(t)~dt~,~\hspace{0in}~\tau \left( t\right) =B\left( t\right) \equiv
\int_{0}^{t}b\left( t^{\prime }\right) dt^{\prime }~,  \label{repar}
\end{equation}%
which maps the metric (\ref{newmetric}) into the Friedmann-Robertson-Walker
metric $g_{\mu \nu }$ with line element
\begin{subequations}
\begin{equation}
ds^{2}=d\tau ^{2}-\alpha (\tau )^{2}~d\vec{x}^{2}~,  \label{FRWmetric1}
\end{equation}%
where
\begin{equation}
\alpha (\tau (t))\equiv a(t)~.  \label{alphadefinition}
\end{equation}

As stated above, our trick to construct \textit{isochronous} dynamical
systems is \textit{formally} based on the eventual replacement of the
physical time variable $t$ with another variable $\tau $. In a general
relativistic context this corresponds to have a \textit{periodic} lapse
function (with an \textit{\ arbitrarily assigned} period $T$) having
moreover a vanishing mean value, so that its integral $B\left( t\right) ,$
see (\ref{repar}), is also periodic with period $T$:
\end{subequations}
\begin{equation}
b(t+T)=b(t)~;~~~\tau \left( t+T\right) =\tau \left( t\right) ~.
\label{bBperiodic}
\end{equation}

The choice (\ref{bBperiodic})---when inserted in (\ref{newmetric})---implies
the following consequences \cite{isochronous cosmologies}.

(i) For any FRW metric (\ref{FRWmetric1}), the corresponding---via the
relation (\ref{bBperiodic})---metric (\ref{newmetric}) is periodic in the
time variable $t$, and therefore it is cyclic, defining thereby an \textit{%
isochronous spacetime}.

(ii) Since the lapse function is periodic with period $T$ and has zero mean
value, it must vanish at an infinite set of times $t=t_{n}\equiv t_{0}\pm
nT/2$ with $n=0,1,2,3,...$, implying that the metric (\ref{newmetric}) is
degenerate at these times.

(iii) If the metric (\ref{FRWmetric1}) is a solution of Einstein's equations
(namely, of the FRW equations) corresponding to the stress energy tensor of
a perfect fluid (or a mixture of different perfect fluids) $T_{\mu \nu }=%
\left[ r(\tau )+p(\tau )\right] u_{\mu }u_{\nu }-g_{\mu \nu }p(\tau )$,
where $r(\tau )$ and $p(\tau )$ are the energy density and pressure of the
fluid and $u_{\mu }=\delta _{\mu 0}$ is its 4-velocity, then (\ref{newmetric}%
) is also a solution of Einstein's equations corresponding to a stress
energy tensor $T_{\mu \nu }=\left[ \rho (t)+P(t)\right] U_{\mu }U_{\nu
}-g_{\mu \nu }P(\tau )$ of a perfect fluid with energy density $\rho
(t)=r(\tau (t))$, pressure $P(t)=p(\tau (t))$ and 4-velocity $U_{\mu
}=|b(t)|\delta _{\mu 0}$. Therefore the metrics (\ref{newmetric}) and (\ref%
{FRWmetric1}) describe two universes filled with the same perfect fluid(s).

(iv) Since for $t\neq t_{n}$ the two metrics (\ref{newmetric}) and (\ref%
{FRWmetric1}) are connected by a diffeomorphic transformation, they give the
same physics in any time interval within which $b(t)$ does \textit{not}
vanish, so that they are in fact experimentally indistinguishable there.

(v) One can conveniently choose the lapse function in such a way that the
scale factor is always positive, i.e. $a(t)=\alpha (\tau (t))>0$ for any $t$%
. This choice implies that all the scalar invariants as well as the pressure
and energy density of the fluid remain finite at any $t$ and the spacetime
is geodesically complete, thus singularity free.

(vi) Geodesics have, in spacetime, a helical structure.

These solutions of course violate the "equivalence principle" at the
infinite set of discrete times $t_{n}.$ It is thus seen that, with the
choice (\ref{bBperiodic}), the class of spacetime solutions of the equations
of general relativity is enlarged by allowing unobservable violations of the
"equivalence principle" at a discrete, numerable set of times $t_{n}$. This
more general class of spacetimes do not seem to entail any \textit{%
experimentally observable} differences; although they might entail a quite
different structure of spacetime, for instance absence versus presence of
Bing Bang singularities.

The problem of finding a generalization of general relativity which includes
degenerate metrics has been studied in the past, motivated by the
consideration of a class of metrics, the so called signature-changing
metrics \cite%
{ellis1,ellis2,elliscarfora,elliscomment,ellis3,ellis4,ellis5,ellis6,ellis8,ellis9,ellis10,ellis11,ellis12,ellis13,ellis14,ellis15,ellis16,ellis17,ellis18,ellis19}
, which give a classical realization of the change of signature in quantum
cosmology conjectured by Hartle and Hawking \cite%
{hartlehawking3,hartlehawking1,hartlehawking2,hartlehawking4,hawkinghistory}%
. The classical change of signature for a homogeneous, isotropic and
spatially flat universe is realized by a metric tensor defined by the
following line element%
\begin{equation}
ds^{2}=N(t)~dt^{2}-R(t)^{2}~d\vec{x}^{2}~,
\label{changeofsignaturedegenerate}
\end{equation}%
where $N(t)$ is a continuous function changing sign at some time $t_{0}$,
for instance $N(t)$ is positive for $t>t_{0}$, negative for $t<t_{0}$ and
zero for $t=t_{0}$ in correspondence to the hypersurface of change of
signature where the metric (\ref{changeofsignaturedegenerate}) is degenerate
(note the difference from (\ref{newmetric}), where the coefficient $%
b^{2}\left( t\right) $ of $dt^{2}$ might vanish at some points but never
becomes \textit{negative, }$b^{2}\left( t\right) \geq 0$).

These generalizations of standard general relativity including degenerate
metrics as (\ref{newmetric}) and (\ref{changeofsignaturedegenerate}) are not
unique since they depend on their defining prescriptions: \textit{different}
prescriptions yield \textit{different} physical theories \cite{elliscomment}%
. In particular, it has been pointed out that different generalizations
based on different prescriptions give different junction conditions on the
degeneracy hypersurfaces, see for instance the discussion in \cite%
{elliscarfora,elliscomment}. In fact "one can also consider discontinuities
in the extrinsic curvature" \cite{elliscarfora} on the degeneracy
hypersurface of a degenerate solution of Einstein's equations, "but attempts
to relate these to a distributional matter source at the boundary require
some form of field equations valid on the surface", see discussion in \cite%
{elliscomment}, and this would require some special prescription on what
Einstein's equations are on the degeneracy hypersurfaces. Thus, if one
accepts degenerate solutions of Einstein's equations as physically
acceptable spacetimes, there is no reason to infer that a discontinuity of
the extrinsic curvature on a degeneracy surface implies that the stress
energy tensor associated to such a metric tensor must have a distributional
form there. What is more, delta functions with support on degeneracy
hypersurfaces have no significant effect in covariant integrals \cite%
{ellis15,ellis19}, due to the fact that the covariant volume element $dx^{4}%
\sqrt{-g}$ associated to a degenerate metric is null on its degeneracy
hypersurface, because the determinant $g$ vanishes there. For instance, for
the metric (\ref{newmetric}) one has, for any integrable function $f\left(
x\right) $,%
\begin{equation}
\int dx^{4}\sqrt{-g}\delta (t-t_{n})f\left( x\right) =\int
dtdx^{3}|b(t)||a(t)|^{3}\delta (t-t_{n})f\left( x\right)
=|b(t_{n})|~|a(t_{n})|^{3}\int \,dx^{3}f\left( x\right) =0
\end{equation}%
(because $b\left( t_{n}\right) =0$). Hence delta-like distributions centered
on the degeneracy hypersurfaces $t=t_{n}$ disappear from integrals (see also
the discussion in \cite{ellis15}).

So, the question of how to include degenerate metrics in general relativity
is moot.

In order to formulate a generalized gravitational theory which includes such
metrics, it is convenient to focus directly on the Einstein's equations
characterizing the geometry of spacetime, which themselves admit degenerate
metrics as their solutions. If, however, one considers desirable to derive
this theory from a variational principle, a possibility is to assume the
standard Einstein-Hilbert gravitational action of general relativity, such
that the total action of gravitation plus the matter (nongravitational)
fields is%
\begin{equation}
S_{Tot}=\int_{M}\sqrt{-g}\left( -\frac{1}{2\mathit{k}}R+\mathit{L}%
_{Mat}\right) d^{4}x~,  \label{actiontotal}
\end{equation}%
where $\mathit{k}=8\pi G/c^{4}$, $G$ is the gravitational constant, $c$ the
speed of light, $M$ is the manifold defining the spacetime and $\mathit{L}%
_{Mat}$ is the Lagrangian density of matter fields; and add the requirement
that, for a metric tensor which is degenerate on some hypersurface $\Sigma $%
, the variation of the metric $g_{\mu \nu }$ vanish on $\Sigma $, i.e. $%
\delta g_{\mu \nu }|_{\Sigma }=0$. With such an assumption the variation of
the action (\ref{actiontotal}) is%
\begin{equation}
\delta S_{Tot}=\frac{1}{2}\int_{M/\Sigma }\sqrt{-g}\left( -\frac{1}{\mathit{k%
}}G^{\mu \nu }+T^{\mu \nu }\right) d^{4}x~,  \label{actiontotalvariation}
\end{equation}%
where $G^{\mu \nu }$ is the Einstein tensor and $T^{\mu \nu }$ is the
stress-energy tensor of matter fields, which gives the standard equations of
general relativity in the region where the metric tensor is nondegenerate.

To bypass the problems related to degenerate signature-changing metrics, it
has been proposed \cite{ellis1,ellis2} to consider a different,
nondegenerate but discontinuous, realization of the classical change of
signature, as given by the metric%
\begin{equation}
ds^{2}=f(\tau )~d\tau ^{2}-R(\tau )^{2}~d\vec{x}^{2},
\label{changeofsignaturediscontinuous}
\end{equation}%
with a discontinuous lapse function $f(\tau )=1$ for $\tau >\tau _{0}$ and $%
f(\tau )=-1$ for $\tau <\tau _{0}$. It has been shown \cite{ellis1,ellis2}
that in this case it is possible to introduce a smooth (generalized)
orthonormal reference frame which allows a variational derivation of
Einstein's equations and an extension of the Darmois formalism \cite{ellis16}
to discontinuous metrics, in such a way that, also in the case of a
discontinuous metric tensor, the discontinuity of the extrinsic curvature on
a hypersurface is related to a distributional stress energy tensor \cite%
{ellis15}. This is achieved by adding a surface term to the Einstein-Hilbert
action, in such a way that the gravitational action becomes%
\begin{equation}
S_{g}=\int_{M/\Sigma }\sqrt{-g}Rd^{4}x+\oint_{\Sigma }\sqrt{-g}Kd\Sigma
\label{actionboundary}
\end{equation}%
where $M$ is the manifold defining the spacetime, $\Sigma $ is the boundary
of $M$ where the metric tensor is discontinuous, $R$ is the Ricci scalar
curvature and $K$ the extrinsic curvature of $\Sigma $. The addition of such
a boundary term is always necessary when one considers manifolds with
boundaries \cite{HawkingHorowitz}. Incidentally we note that this action
cannot be used in the case of degenerate metrics, because in this case the
degeneracy surface $\Sigma $ has no unitary normal vector hence the
extrinsic curvature $K$ does not exist.

The two realizations of the classical change of signature, the continuous
and degenerate one, see (\ref{changeofsignaturedegenerate}), and the
discontinuous one, see (\ref{changeofsignaturediscontinuous}), are \textit{%
locally but not globally} diffeomorphic, since they are related by a change
of the time variable $dt=d\tau /\sqrt{|N(t)|}$ which is not defined at the
time of signature change $t_{0}$ when $N(t)=0$. Hence they represent two
different spacetimes. However, except for the instants when $N(t)$ vanishes,
they are \textit{locally } diffeomorphic and thus they describe---excepts at
those instants---the same physics .

As in the case of signature-changing metrics, it is also possible to
consider a realization of \textit{isochronous} cosmologies different from
that we introduced in \cite{isochronous cosmologies} and reviewed above (see
(\ref{newmetric}) and the subsequent treatment), via \textit{nondegenerate}
metrics featuring a finite jump of their first derivatives at an infinite,
discrete set of equispaced times. Such a realization is given by the
following metric:%
\begin{equation}
ds^{2}=d\eta ^{2}-\tilde{a}(\eta )^{2}d\vec{x}^{2}
\label{newmetricnondegenerate}
\end{equation}%
with
\begin{equation}
\alpha (\tau (\eta ))\equiv \tilde{a}(\eta )
\label{newscalarfactornondegenerate}
\end{equation}%
and
\begin{equation}
d\tau =C(\eta )~d\eta ,  \label{newCndegenerate}
\end{equation}%
with a periodic function $C(\eta )$ of period $T$ given for instance by $%
C(\eta )=1$ for $nT<\eta <(2n+1)T/2$ and $C(\eta )=-1$ for $(2n+1)T/2<\eta
<(n+1)T$ and $n$ integer, hence such that
\begin{subequations}
\begin{equation}
\tau \left( \eta \right) \equiv \int_{0}^{\eta }C\left( \eta ^{\prime
}\right) d\eta ^{\prime }=\eta -nT~~~\text{for}~~~nT<t<\left( 2n+1\right)
T/2~,
\end{equation}%
\begin{equation}
\tau \left( \eta \right) \equiv \int_{0}^{\eta }C\left( \eta ^{\prime
}\right) d\eta ^{\prime }=\left( n+1\right) T-\eta ~~~\text{for}~~~\left(
2n+1\right) T/2<t<\left( n+1\right) T~.
\end{equation}

The metric (\ref{newmetricnondegenerate}) is periodic with period $T$ and is
not degenerate at the inversion times $\eta _{n}=nT/2$, where its first
derivatives feature a finite jump. Since (\ref{newmetricnondegenerate}) is
nondegenerate, now the Darmois formalism \cite{ellis16} applies and one can
relate the discontinuity of the first derivatives of the metric, hence the
discontinuity of the extrinsic curvature of the hypersurfaces $\Sigma _{\eta
_{n}}$ defined by $\eta =\eta _{n}$ (which is now well defined), with a
distributional stress-energy tensor. In fact defining the unitary normal
vector $n_{\alpha }=\delta _{\alpha 0}$, the tangent vectors $e_{\alpha
}^{a}=\delta _{\alpha }^{a}$ and the induced metric $h_{ab}=-\tilde{a}(\eta
)^{2}\delta _{ab}$ of a hypersurface $\Sigma _{\eta }$ of equation $\eta
=constant$, which also implies $n^{\alpha }=\delta ^{\alpha 0}$, $%
e_{a}^{\alpha }=\delta _{a}^{\alpha }$ and $h^{ab}=-\delta _{ab}/\tilde{a}%
(\eta )^{2}$, the extrinsic curvature of $\Sigma _{\eta }$ is given by \cite%
{poisson}
\end{subequations}
\begin{equation}
K_{ab}\equiv n_{\alpha ;\beta }e_{a}^{\alpha }e_{b}^{\beta }= \alpha (\tau
(\eta )) \, \alpha^{\prime }(\tau (\eta )) \, C(\eta ) \,\delta _{ab}~,
\label{extrinsiccurvaturenondegenerate}
\end{equation}%
where $\alpha^{\prime }(z)\equiv d\alpha(z)/dz$ \footnote{%
Note that in \cite{isochronous cosmologies} we have calculated the extrinsic
curvature of the hypersurfaces $t=const$ of the metric (\ref{newmetric})
defining the normal vector of such hypersurfaces as $n^{\alpha }=\delta
^{\alpha }/b(t)$, the tangent vectors as $e_{\alpha }^{a}=a(t)\delta
_{\alpha }^{a}$ and $h_{ab}=-\delta _{ab}$. Instead, (\ref%
{extrinsiccurvatureJUMPnondegenerate}) corresponds to the choice $n^{\alpha
}=\delta ^{\alpha 0}/|b(t)|$, $e_{\alpha }^{a}=\delta _{\alpha }^{a}$ and $%
h_{ab}=-a(t)^{2}\delta _{ab}$.}.

Therefore the discontinuity of the extrinsic curvature on the hypersurface $%
\Sigma _{\eta _{n}}$ is%
\begin{equation}
\lbrack K_{ab}]|_{\Sigma _{\eta _{n}}}=2\,(-1)^{n}\,\alpha (\tau (\eta
))\,\alpha ^{\prime }(\tau (\eta ))\,\delta _{ab}~,
\label{extrinsiccurvatureJUMPnondegenerate}
\end{equation}%
and this implies that the stress-energy tensor has, on the numerable set of
hypersurfaces $\Sigma _{\eta _{n}}$, a distributional contribution given by%
\begin{equation}
T_{\alpha \beta }^{distr}=\sum_{n}\delta (\eta -\eta _{n})\,S_{ab}|_{\Sigma
_{\eta _{n}}}e_{\alpha }^{a}e_{\beta }^{b}~,
\label{StressEnergyTensorNondegenerate}
\end{equation}%
where%
\begin{equation}
S_{ab}|_{\Sigma _{\eta _{n}}}\equiv \frac{1}{8\pi }\left( [K_{ab}]|_{\Sigma
_{\eta _{n}}}-[K]|_{\Sigma _{\eta _{n}}}h_{ab}\right) =\frac{(-1)^{n+1}}{%
2\pi }\,\alpha (\tau (\eta ))\,\alpha ^{\prime }(\tau (\eta ))\,\delta _{ab}
\end{equation}%
and $[K]|_{\Sigma _{\eta _{n}}}\equiv \lbrack K_{ab}]|_{\Sigma _{\eta
_{n}}}h^{ab}$. We note that this distributional stress-energy tensor is
absent in the case of a vacuum Minkowskian universe, since in this case $%
\alpha (\tau )$ is constant and its derivative vanishes.

Again, let us point out that the two metrics (\ref{newmetric}) and (\ref%
{newmetricnondegenerate}) are two \textit{different} realizations of an
\textit{isochronous} cosmology, since they are \textit{locally}, but \textit{%
not globally}, diffeomorphic via the change of variable $d\eta =b(t)dt$
which is singular at $t_{n}$. In particular the statements (i,iii,iv,v,vi),
see above, which are valid for the metric (\ref{newmetric}), also apply to
the metric (\ref{newmetricnondegenerate}). Moreover, both (\ref{newmetric})
and (\ref{newmetricnondegenerate}) are \textit{locally} (but \textit{not
globally}) diffeomorphic to the FRW metric (\ref{FRWmetric1}) and therefore
they gives an \textit{identical} dynamics in any time interval such that $%
b(t)\neq 0$ or $C(\eta )\neq 0$. The main difference is that the isochronous
evolution given by (\ref{newmetric}) entails the introduction of degenerate
metrics and the nondegenerate realization of the isochronous dynamics given
by (\ref{newmetricnondegenerate}) implies the existence of a distributional
contribution to the stress-energy tensor. In both cases the evolution of the
universe is locally the same as that described by the FRW equations,
implying that it is not possible to discriminate by experiments local in
time between an isochronous and a noncyclic evolution of the universe.

\section{Conclusions}

In this paper we have reviewed and further discussed the idea introduced in
\cite{isochronous cosmologies}, itself being an extension to a general
relativity context---hence relevant to discuss cosmological models---of a
finding valid in the wide context of nonrelativistic Hamiltonian dynamics.
Let us emphasize that---as in that context---the essence of our observation
is to point out the disturbing possibility, given a physical theory that
describes reality, to manufacture variations of it---i. e., different
theories, characterized by the introduction of an arbitrary parameter $T$%
---which describe essentially the \textit{same} reality over time intervals
smaller than $T$ but are \textit{cyclic} with period $T$.

And let us conclude this paper by emphasizing that we are \textit{not}
ourselves arguing that our Universe evolves \textit{cyclically} indeed
\textit{isochronously}, much less wish we to enter the related philosophical
issues (not our cup of tea!); we have merely pointed out the (disturbing!)
fact that, in the context of theoretical and mathematical physics, such a
possibility does not seem to be excluded; nor does it seem feasible to
exclude it by realizable experiments.

\bigskip

\section{Acknowledgements}

This paper was completed while FB was a Marie Curie Fellow of the Istituto
Nazionale di Alta Matematica "Francesco Severi".

\end{document}